\documentclass[prb,twocolumn,showpacs,preprintnumbers,amsmath,amssymb,superscriptaddress]{revtex4}

\usepackage{graphicx}

\usepackage{dcolumn}
\usepackage{bm}

\begin{document}

\title{Complex trend of magnetic order in Fe clusters on 4$d$ transition-metal surfaces}

\author{V. Sessi}
\affiliation{European Synchrotron Radiation Facility, 6 rue Jules Horowitz, BP 220 38043 Grenoble Cedex 9, France}
\author{F. Otte}
\affiliation{Institute of Theoretical Physics and Astrophysics, University of Kiel, Leibnitzstr. 15, 24098 Kiel, Germany}
\author{S. Krotzky}
\affiliation{Max-Planck-Institut f\"ur Festk\"orperforschung,
Heisenbergstrasse 1, 70569 Stuttgart, Germany}
\author{C. Tieg}
\affiliation{European Synchrotron Radiation Facility, 6 rue Jules Horowitz, BP 220 38043 Grenoble Cedex 9, France}
\author{M. Wasniowska}
\affiliation{Max-Planck-Institut f\"ur Festk\"orperforschung,
Heisenbergstrasse 1, 70569 Stuttgart, Germany}
\author{P. Ferriani}
\affiliation{Institute of Theoretical Physics and Astrophysics, University of Kiel, Leibnitzstr. 15, 24098 Kiel, Germany}
\author{S. Heinze}
\affiliation{Institute of Theoretical Physics and Astrophysics, University of Kiel, Leibnitzstr. 15, 24098 Kiel, Germany}
\author{J. Honolka}
\affiliation{Max-Planck-Institut f\"ur Festk\"orperforschung,
Heisenbergstrasse 1, 70569 Stuttgart, Germany}
\affiliation {Institute of Physics of the ASCR v. v. i., Na Slovance 2, 182 21 Prague, Czech Republic}
\author{K. Kern}
\affiliation{Max-Planck-Institut f\"ur Festk\"orperforschung,
Heisenbergstrasse 1, 70569 Stuttgart, Germany}
\affiliation{Institut de Physique de la Mati\`ere Condens\'ee, Ecole Polytechnique F\'ed\'erale de Lausanne, CH-1015 Lausanne, Switzerland}

\date{\today}

\begin{abstract}

We demonstrate the occurrence of compensated spin configurations in Fe clusters and monolayers
on Ru(0001) and Rh(111) by a combination of X-ray magnetic circular dichroism experiments and first-principles calculations.
Our results reveal complex intra-cluster exchange interactions which depend strongly on the
substrate 4$d$-band filling, the cluster geometry as well as lateral and vertical structural relaxations.
The importance of substrate 4$d$-band filling manifests itself also in small nearest-neighbor exchange
interactions in Fe dimers and in an nearly inverted trend of the Ruderman-Kittel-Kasuya-Yosida coupling constants
for Fe adatoms on the Ru and Rh surface.

\end{abstract}


\maketitle

\section{Introduction}

Today, there is a strive for a controlled fabrication of nanomagnets in order to explore the concepts
of spintronics at the atomic scale.
Much progress was achieved in understanding direct intracluster exchange interactions in ferromagnetic
few-atom clusters situated on metal surfaces~(see
e.g.~Ref.[\onlinecite{Enders2010}]
and references therein),
as well as indirect surface-mediated magnetic Ruderman-Kittel-Kasuya-Yosida (RKKY) interactions~\cite{Zhou2010, Khajetoorians2012}. A central challenge remains the increasing importance of thermal fluctuations in few-atom clusters, which leads to unwanted destabilization of moments.
As a consequence, in recent years the research focus has shifted towards heavy $5d$ transition-metal substrates, where large spin-orbit coupling (SOC) gives hope to enhance the magnetic anisotropy and to counteract superparamagnetic behavior.
Indeed, for ferromagnetic Co structures on Pt(111)
experiments show extraordinary large magnetic anisotropies of up to 9~meV/atom~\cite{Gambardella2003, Rusponi2003}.
However, it has been recently realized that in transition-metal
nanostructures on surfaces, SOC also induces the Dzyaloshinskii-Moriya (DM) interaction~\cite{Bode2007}.
It favors non-collinear magnetic configurations and can destabilize ferromagnetism even on the atomic scale~\cite{Ferriani2008, Mankovsky2009}.

Less attention has been given to the lighter 4$d$ transition metal substrates~\cite{Robles2000, Lee2002, Sawada2003, Yokoyama2003, Przybylski2006, 6},
where effects of relativistic origin such as the DM term and the magnetocrystalline anisotropy are expected to be much smaller.
The exchange interaction, on the other hand, can depend critically on the hybridization with the surface and its band filling.
Based on first-principles calculations it has been predicted that the nearest-neighbor (NN) exchange coupling changes
from antiferromagnetic (AFM) to ferromagnetic (FM) for Fe monolayers on Ru(0001) and Rh(111), respectively \cite{3}.
Since its magnitude is small, interactions beyond NNs as well as higher-order terms beyond the pair-wise
Heisenberg exchange, such as the four-spin and biquadratic interactions, can play a decisive role for the magnetic order~\cite{3,3a}.
Magnetic configurations that are
surprising for Fe have been predicted for those substrates, namely a N\'eel state with angles of $120^\circ$ between adjacent spins for
Fe monolayers on Ru(0001), and a collinear double row-wise AFM $uudd$-state on Rh(111).
These two systems are thus ideal candidates to systematically study the formation of complex magnetic phases driven by frustrated interactions beyond NN Heisenberg exchange.

Here we show the essential importance of Fe 3$d$ state itinerancy and hybridization with partially filled 4$d$ substrate bands in monatomic-height
Fe clusters of different atomic size $N$ and various geometries. Randomly positioned single Fe atom spins in the dilute regime ($N=1$) indirectly interact via the RKKY mechanism which shows inverted character on Ru(0001) and Rh(111).
For Fe dimers ($N=2$) we prove the AFM (Ru) to FM (Rh) cross-over of the NN exchange coupling constant $J_1$,
and for larger clusters ($2<N\leq4$) the onset of cluster geometry dependent compensated magnetic structures, both predicted by our first-principles calculations. We demonstrate that compact clusters are ferromagnetic while open structures exhibit compensated antiferromagnetic states. The origin of this unexpected trend arises from the competition of direct Fe-Fe exchange in the clusters
and indirect exchange mediated by the substrate.
Finally, we present experimental evidence for the formation of compensated spin textures both for Ru(0001) and Rh(111)
in fully ordered epitaxial Fe islands.

\section{Experimental}
\begin{figure*}
\centering
\includegraphics[width=1.0\textwidth]{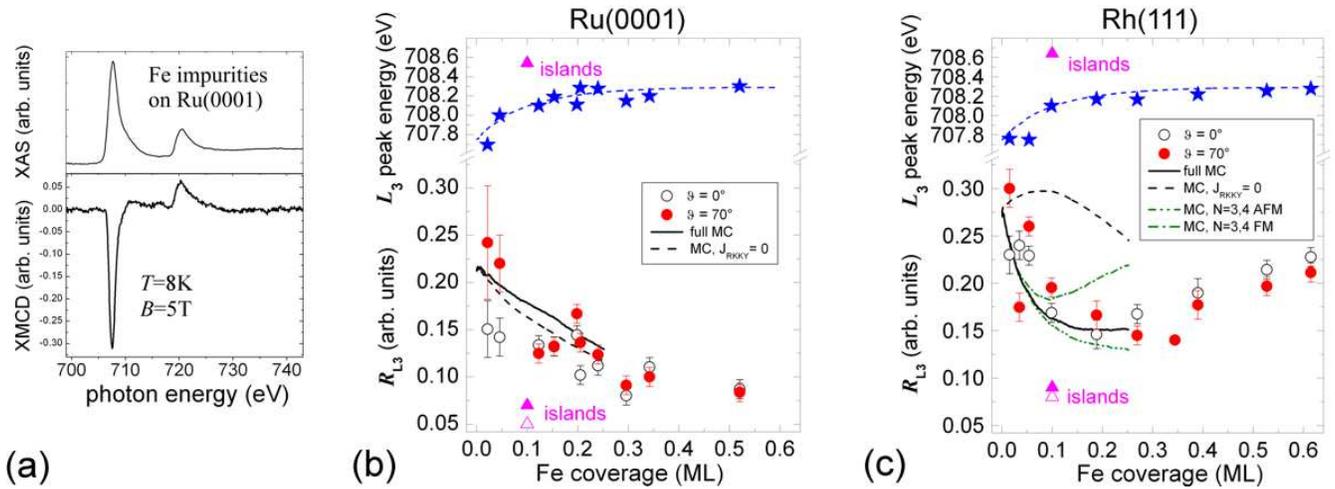}
\caption{Quench-condensed deposition of Fe on Ru(0001) and Rh(111). (a) Examples of XAS and XMCD spectra in the impurity limit. (b) and (c): {\underline{Top}} Measured XAS $L_3$ peak photon energy (stars). The dashed line is a guide to the eye.
{\underline{Bottom}} Average Fe magnetization $R_{L_3}$ versus Fe coverage. The full and dashed black lines are combined MC simulations with and without $J_{\text{RKKY}}$ between single adatoms, respectively. For comparison, dash-dotted curves represent simulations assuming FM and AFM trimers and tetramers.
Triangles in magenta show values measured on epitaxial islands at $\theta= 0.1$~ML.}
\label{impurities}
\end{figure*}
X-ray magnetic circular dichroism (XMCD) experiments were carried out at the ID08 beamline of the European Synchrotron Radiation Facility (ESRF), where samples can be prepared in-situ under UHV conditions.\newline
Ru(0001) and Rh(111) single crystal surfaces were prepared by cycles of Ar-sputtering and annealing at 900$^{\circ}$C. A scanning tunneling microscope allows to verify the cleanliness of the single crystal surfaces.
Fe of 99.99$\%$ purity was deposited onto Ru(0001) and Rh(111) from a rod by electron bombardment heating. A precise calibration of the evaporator was done with the help of the scanning tunneling microscope.\newline
During the X-ray measurements the pressure in the magnet chamber was $<3\times 10^{-10}$mbar. Possible contaminations containing oxygen were excluded by monitoring the O K-edge signal at around $540$eV.
X-ray absorption spectra (XAS) were measured at the Fe $L_{3,2}$-edges with 99$\%$ positive and negative circularly polarized light ($\sigma^{+}$ and $\sigma^{-}$) using the surface sensitive total electron yield (TEY) mode. XMCD and XAS signals are then defined as the difference $(\sigma^{+}-\sigma^{-})$ and the average $(\sigma^{+}+\sigma^{-})/2$, respectively. The Fe $L_{3,2}$ XAS contribution to the TEY is obtained by subtraction of a background signal measured prior to Fe deposition.
Spectroscopy was done at two angles of incidence with respect to the sample surface: $\vartheta=70^{\circ}$ (in-plane) and $\vartheta=0^{\circ}$ (polar). Magnetic fields up to $B=5$~T are applied parallel to the X-ray beam direction.
Both XAS and XMCD signals scale with the iron coverage $\theta$.
Thus, all XMCD data shown in this work are normalized to the respective $L_3$ peak amplitude in the non-dichroic Fe XAS. The $L_3$ peak value $R_{L_3}$ of the normalized XMCD is then a good measure of the projection of the average magnetization $<\mathbf{M}>$ on the field direction $\hat{z}$: $R_{L_3}\sim \mathbf{P}_{\hat{z}}\cdot<\mathbf{M}>$.

\section{Results and Discussion}
Using {\it in-situ} quench-condensed deposition of submonolayer amounts of Fe at low temperatures we achieve
a statistical distribution $\Gamma(N,g)$ of cluster sizes $N$ and their respective geometries $g$ on both Rh(111) and
Ru(0001) due to suppression of diffusion of surface adatoms.
Fig.~\ref{impurities}(a) shows examples of XAS and XMCD spectra of impurities measured at $B=5$T and $T=8$K. A sharp, atomic-like dichroic signal corresponding to $R_{L_3}\sim 0.25$ is visible, as expected for a non-saturated, thermally fluctuating single Fe atom spin moment.
For comparison, saturated Fe atoms on Pt(997) give enhanced values of 0.6 under similar conditions~\cite{5a}.

The impact of an increasing Fe coverage, and thus average Fe-Fe coordination $n_{\text{Fe-Fe}}$, is summarized in Fig.~\ref{impurities}(b) and (c) for Ru(0001) and Rh(111), respectively.
First we note that the non-dichroic XAS $L_3$ peak positions shift by $\Delta=0.6$eV to higher photon energies (blue data). Positive shifts are characteristic for increasing hybridization between Fe 3$d$ states, which leads to more efficient screening of core-hole effects during X-ray absorption. The function $\Delta(n_{\text{Fe-Fe}})$ is usually highly non-linear, saturating already at small values $n_{\text{Fe-Fe}}$~\cite{Hirsch}.

At the bottom parts of Fig.~\ref{impurities}(b) and (c) the evolution of the magnetic signal with raising $n_{\text{Fe-Fe}}$ is shown.
The trend of $R_{L_3}(\theta)$ for Fe on Ru(0001) shows a steady decay of the average magnetization with Fe coverage, indicating progressive magnetic compensation. Comparing the $R_{L_3}$ values for $\vartheta=70^{\circ}$ and $\vartheta=0^{\circ}$, we observe an in-plane magnetic anisotropy in the range $\theta<0.1$~ML. For Fe clusters on Rh(111) we find
an even steeper initial decrease of $R_{L_3}$ with coverage and at low coverages an in-plane magnetic anisotropy.
In contrast to Ru(0001), at an intermediate coverage $\theta_m \sim 0.25$~ML the magnetization reaches a minimum and increases monotonously thereafter.

In order to understand the observed trends of the magnetization with coverage, we have performed first-principles
calculations based on density functional theory (DFT) for Fe clusters of different size and shape on both surfaces. We have applied
the projector-augmented-wave method as implemented in the Vienna Ab-Initio Simulation Package (VASP) \cite{VASP1,VASP2} and
used the generalized gradient approximation (GGA) to the exchange-correlation functional \cite{PBE}.
We consider different collinear magnetic configurations of the clusters and compare their total energies
taking vertical and lateral structural relaxations into account.
Computational details can be found in the Appendix A.

\begin{figure}
\centering
\includegraphics[width=0.40\textwidth]{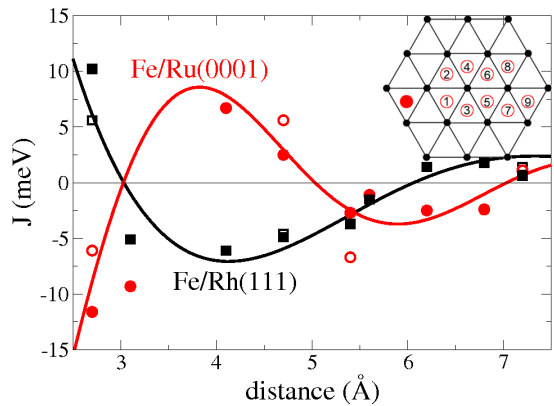}
\caption{Calculated exchange coupling constants $J(r)$ as a function of separation $r$ between pairs of Fe adatoms.
Open symbols denote results for pure hcp adsorption sites and filled symbols mark pure fcc sites and mixed dimers.
The inset shows the position of the Fe atoms in the dimer. The fitting has been performed with an RKKY-like function.
}
\label{exchange_J}
\end{figure}

At low coverage there will be mostly a distribution of single adatoms which can interact with each other via the exchange interaction
mediated by the substrate. Therefore, we first focus on the exchange interaction between two Fe adatoms as a function of their
distance. The exchange constants $J(r)$ obtained from total energy calculations are presented in Fig.~\ref{exchange_J}. As expected
we observe an oscillatory behavior of $J(r)$ changing from FM ($J>0$) to AFM ($J<0$) and a decay of its magnitude
with increasing Fe-Fe separation. Interestingly, the trend found for Fe dimers on the Rh and Ru surface is almost perfectly inverted.
In contrast to the exchange interaction reported for substrates with a filled $d$-band \cite{Zhou2010,FeCu111,FePd,Mavro}
we find that the NN exchange constant $J_1$
is reduced by about one order of magnitude and, thus, in competition with indirect exchange interactions $J_n$ with $n>1$.
The latter will in the following be referred to as $J_{\text{RKKY}}$.

\begin{figure}
\centering
\includegraphics[width=0.45\textwidth]{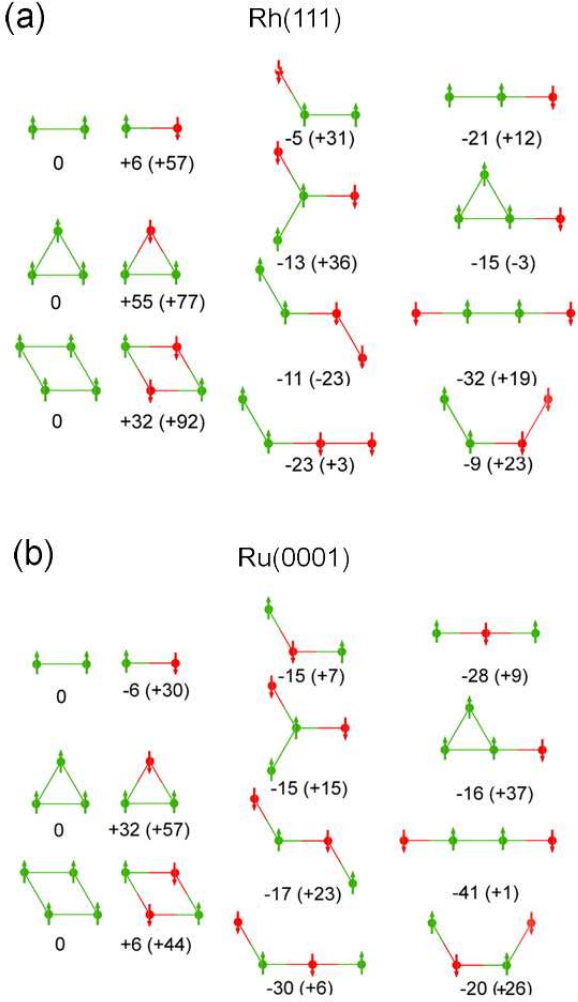}
\caption{Total energy differences between different magnetic configurations for the Fe dimer, trimers and tetramers on
(a) Rh(111) and (b) Ru(0001).
Energy differences in meV per Fe atom
are given with respect to the FM state.
Values in brackets are energy differences without taking structural relaxations into account.}
\label{dE_clusters}
\end{figure}

Our calculations for Fe trimers and tetramers on
Rh(111) and Ru(0001), shown in Fig.~\ref{dE_clusters}(a) and (b), respectively,
display a complex dependence of the magnetic order on the cluster geometry.
For Rh(111) we find that compact trimers and tetramers possess a FM ground state which is in accordance with
the FM NN exchange coupling from the dimer calculations (cf.~Fig.~\ref{exchange_J})
although the energy differences are much larger than expected from the exchange constants obtained from the dimers.
However, Fe clusters in an open structure show a tendency to AFM order with compensated spin structures. This is surprising in view of the
FM exchange interaction of the dimers. Interestingly, the open tetramers already display the $uudd$ state predicted for the full monolayer.

These effects arise due to a competition of direct Fe-Fe exchange and indirect exchange mediated by the substrate
which are closely linked with the cluster geometry and structural relaxations that differ for open and compact
structures~\cite{Fabian}.
The impact of the structural relaxation on the magnetic state is evident from
Fig.~\ref{dE_clusters}(a)
if one compares the energy differences obtained without taking structural relaxations into account. In the case of
a NN dimer on Rh(111) the exchange energy is reduced by one order of magnitude upon relaxation, leading to the very
low value of 6~meV/Fe-atom.
A considerable reduction of the energy difference occurs also for the compact trimers and tetramers.
For most of the open cluster configurations, the energetically favorable state even changes from a FM to a
compensated state upon taking structural relaxations into account.

A similar trend of magnetic order is found for Fe trimers and tetramers on Ru(0001) as shown in
Fig.~\ref{dE_clusters}(b).
For open structures, compensated AFM spin structures are found which is expected
from the AFM NN exchange in Fe dimers (cf.~Fig.~\ref{exchange_J}).
Note, that the AFM NN exchange is driven by the hybridization with the substrate. This can be seen by comparing the energy
differences for the Fe dimer without structural relaxation which prefers a FM state (see Fig.~\ref{dE_clusters}(b)).
However, compact trimers and tetramers are in a FM ground state. The origin of this unexpected change of exchange
coupling in the clusters is due to the enhanced direct FM Fe-Fe exchange interaction and a weakened
effect of the Ru substrate.

In order to obtain a quantitative interpretation of our experimental data based on the magnetic configurations calculated
from first-principles we performed Monte-Carlo (MC) simulations. Knowing the magnetic ground states of all cluster configurations $(N,g)$ with $N\leq4$, MC simulations of $\Gamma(N,g)$ allows us to estimate the coverage dependent average magnetization of the ensemble in a magnetic field $B=5$T. 
We assume each cluster to be magnetically independent and that each single adatom interacts only with one closest single atom via the RKKY interaction as described in the supplementary.
The magnetic contribution of a certain cluster with $(N,g)$ to the total signal $R_{L_3}$ is then given by a Boltzmann statistics weighted according to $\Gamma(N,g)$, where also induced substrate moments enter the Zeeman energy term
(see Appendix B for further details).

The low coverage behavior ($\theta < 0.1$ML) of $R_{L_3}$ shown in Fig.~\ref{impurities}
can be understood based on the RKKY interactions and the NN exchange constant $J_1$. In the simpler
case of Ru(0001) the magnetization trend at low coverages is dominated by the AFM NN exchange constant $J_1<0$.
In Fig.~\ref{impurities}(b) the result for magnetically independent clusters excluding RKKY interactions is shown, which reproduces the continuous decay of the magnetization well, considering that the modeling contains no free parameter. At lowest coverage, $R_{L_3}$ corresponds to a single spin moment of 3.0$\mu_{\text B}$ as obtained from our DFT calculations in the corresponding Zeeman field.

Turning to the case of Fe clusters on Rh(111), we find that $R_{L_3}$ at lowest coverages is larger compared to the values found
for Ru(0001), which we attribute to (i) the enhanced spin moment 3.2$\mu_{\text B}$ of a single Fe atom on Rh(111) and (ii) the about ten times larger magnetic susceptibility of Rh(111) leading to larger induced substrate moments. The latter enter the Boltzmann statistics via the Zeeman term and stabilize the Fe spin moments. It is evident that even a qualitative understanding of the trend $R_{L_3}(\theta)$ based on the NN exchange interaction is impossible in the case of Rh. The steep
decrease of $R_{L_3}$ at lowest coverages is surprising in view of the positive NN exchange coupling $J_1$.
Starting from single atoms the increase of $\theta$ should thus enhance the average magnetization per Fe atom due to FM dimer formation as seen by the dashed curve in Fig.~\ref{impurities}(c).
However, if we take into account the RKKY coupling between single Fe atoms on Rh(111) we observe that the AFM
exchange coupling for separations of up to 6~${\text{\AA}}$ overcompensate by far the contribution of the FM NN dimer coupling
and accurately reproduces the steep decrease of the average magnetization below $\theta=0.1$ML (solid curve in Fig.~\ref{impurities}(c)).

At intermediate coverages, mostly the formation of FM dimers and FM compact trimers on Rh(111) leads to a plateau in $R_{L_3}$ in good agreement with our experimental data. According to statistics, trimer configurations start to play a role at coverages $\theta>0.1$ML, which again suppress the average moment due to intrinsic compensated structures (see Fig.~\ref{dE_clusters}).
For the excellent quantitative agreement between experiment and simulation, the geometry dependent ground states of tetramers
as obtained from DFT are nevertheless important. This is visible in the two simulations shown in Fig.~\ref{impurities}(c)
for comparison in which it has been assumed that all trimers and tetramers are either perfectly FM or in a maximum compensated
magnetic state.

Our simulations are valid up to coverages of about $0.3$~ML. Beyond that the simulated values $R_{L_3}$ start to decrease due to the increasing spectral weight of clusters with $N>4$, which in our MC simulations are assumed to have zero moment (see Appendix B). We attribute the rise of the experimental $R_{L_3}$ signal to the formation of three-dimensional FM clusters which are less coupled to the substrate and thus will be dominated by the FM direct exchange between Fe moments.


\begin{figure}
\centering
\includegraphics[width=0.5\textwidth]{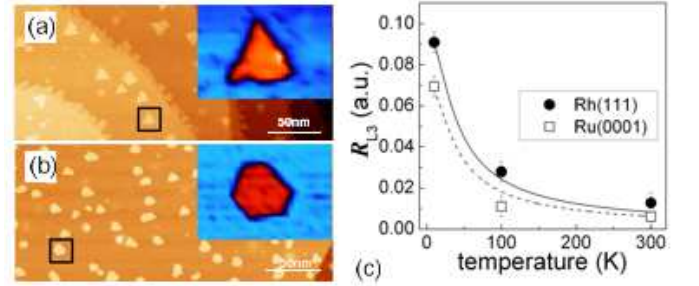}
\caption{(a) and (b): STM topographies of Fe islands on Ru(0001) and Rh(111), respectively. (c) $R_{L_3}$ vs temperature measured at $B=5$T and $\vartheta=70^{\circ}$. Full and dotted lines are Boltzmann statistics of a superparamagnetic macrospin $\mathcal{M}_{\text{Fe}}$.}
\label{islands}
\end{figure}

Finally, we present experimental evidence for compensated magnetic ground states of Fe MLs on the hexagonal surfaces Ru(0001) and
Rh(111).
From the data discussed so far only
the measurements on Ru(0001) are compatible with such a compensated ground state, since the values for $R_{L_3}$ reach very low values at high coverages $\theta=0.5$ML (see Fig.~\ref{impurities}(b)).
For Rh(111) it is evident that such a state cannot be reached by quench-condensed deposition. This is not really surprising since the structure is expected to be disordered, and beyond NN corrections are strongly hampered in a random fashion. We therefore test our systems in the presence of structural order.

In Fig.~\ref{islands}~(a)-(b) STM topographies of 5-10nm wide monatomic-height Fe islands on Ru(0001) and Rh(111) are shown, respectively, which grow epitaxially at deposition temperatures of $T=300$K. On Ru(0001) triangular shaped islands with 5-10nm in diameter are formed on the terraces, and smaller islands decorate the terrace step edges. The onset of the 2nd layer formation on the islands is only visible on Ru(0001), but the ratio between bilayer and monolayer areas corresponds to less than $5\%$. On Rh(111), islands of mostly truncated triangular shape are randomly distributed.

Fig.~\ref{islands}(c) shows $R_{L_3}$ for the two systems measured at $B=5$T and different temperatures. At $T=8$~K only a small Fe dichroic signal of $R_{L_3}=0.07$ and $R_{L_3}=0.09$ is present for $\theta=0.1$ML on Ru(0001) and Rh(111), indicative of intrinsically compensated magnetic ground states in both cases. The temperature dependence up to $T=300$~K can be fitted by a classical Boltzmann statistics of a constant superparamagnetic macrospin $\mathcal{M}_{\text{Fe}}$~\cite{superparamagnetism}, suggesting stable ground states up to energy scales beyond 25meV.
As in the quench-condensed samples a faint in-plane magnetic easy direction is observed for both substrates (see Fig.~\ref{impurities}(b) and (c), where open/full triangles correspond to $\vartheta=0^{\circ}$ and $\vartheta=70^{\circ}$).

The difference of the island results compared to those obtained by quench-condensed deposition underlines the importance of the
structure on the magnetic state. The stabilization of a compensated magnetic configuration on Rh(111) against a FM exchange term
$J_1>0$ is only possible for ordered compact clusters, which allow effective hybridization of Fe 3$d$ states over larger distances. Increased hybridization in the ordered case is also directly visible in the measured XAS $L_3$ peak photon energy, which remains $\sim0.4$eV above the value of quench-condensed structures at largest coverages (see Fig.~\ref{impurities}(b) and (c), top).

\section{Conclusions}
In conclusion, we have shown a complex trend of magnetic order in Fe nanostructures on $4d$ transition-metal surfaces
due to the hybridization of Fe $3d$-states with the partly filled substrate $4d$-band. For Fe dimers the nearest-neighbor
exchange is very small and of opposite sign on the Ru and Rh surface. For larger clusters the competition of direct FM
Fe-Fe exchange with the indirect exchange mediated by the substrate determines the magnetic order.
Finally, we have presented first experimental evidence for the formation of compensated spin textures in epitaxial Fe islands on both for Ru(0001) and Rh(111) as predicted by theory.

\begin{acknowledgements}
F.~O., P.~F., and S.~H. acknowledge financial support by the Deutsche Forschungsgemeinschaft within the SFB 677
and thank the HLRN for providing high-performance computing resources. J.~H. acknowledges the Czech Purkyn\v{e} fellowship program.
\end{acknowledgements}

\appendix
\section{Computational details}\label{DFT}
Fe clusters on Rh(111) and Ru(0001) have been studied based on density functional theory
calculations in the generalized gradient approximation (GGA) to the
exchange-correlation functional~\citep{PBE}, using the projecter-augmented-wave method as implemented in the Vienna Ab-Initio Simulation Package (VASP)~\cite{VASP1,VASP2,VASP3,VASP4,VASPPAW}.
All calculations have been performed in the scalar-relativistic approximation, i.e.
neglecting the effect of spin-orbit coupling.
To model the Fe clusters we have used the $p(4\times4){}$ and $p(5\times5){}$ surface unit cells
for clusters with $N<4{}$ and $N=4$, respectively.
To model the Rh(111) or Ru(0001) surface eight layers have been used. The adatoms as well as the two upmost surface layers have been
structurally relaxed until the forces were smaller than 0.005 eV/\AA. A ($5\times 5 \times 1$) and  ($3\times 3 \times 1$) $\Gamma$-centered
k-point mesh has been used for the $p(4\times4){}$ and  $p(5\times5){}$ surface unit cell, respectively.
The experimental lattice constant of 3.8034~{\AA} for Rh and lattice parameters of
2.7059~{\AA} and 4.2815~{\AA} for Ru have been chosen. The energy cutoff parameter for the plane wave expansion
was 390 eV and a Gaussian smearing of $\sigma=0.07{}$ eV has been applied.

An important aspect of our approach to calculate the exchange constants is the possible
interaction of the clusters with those in adjacent cells
due to the two-dimensional (2D) periodic boundary conditions. In order to estimate the influence of atoms in
neighboring cells we have performed test calculations in the $p(3\times3)$, $p(4\times4)$, and $p(5\times5){}$ unit
cells. We found that the $p(4\times4){}$ unit cell size is sufficiently large to avoid spurious interaction effects
for compact clusters and to determine their magnetic ground state.
However, if the distance between the adatoms within the unit cell is large
as for dimers with large separation,
the influence of atoms in the adjacent unit cells also becomes important. We have taken such interactions into account
when determining the RKKY exchange constants.
More detailed information can be found in Ref.~[\onlinecite{Fabian}].


\section{Monte Carlo Simulations}\label{MC}
During the Monte Carlo (MC) simulation iron atoms are randomly deposited onto two hexagonal hcp and fcc sublattices, each of which have a size of 500$\times$500. Since in the experiment the atoms are deposited at a temperature of $T=8$~K no thermal activated hopping of atoms is included in the simulation. However, we do take into account random tip-over processes onto neighboring free adsorption sites, if the initial MC step chooses a landing site which is already occupied by an iron atom. During one MC deposition cycle the sum of the number of atoms on both sublattices is increased by 0.02$\%$ of a full monolayer (ML).

\begin{figure}
\centering
\includegraphics[width=0.45\textwidth]{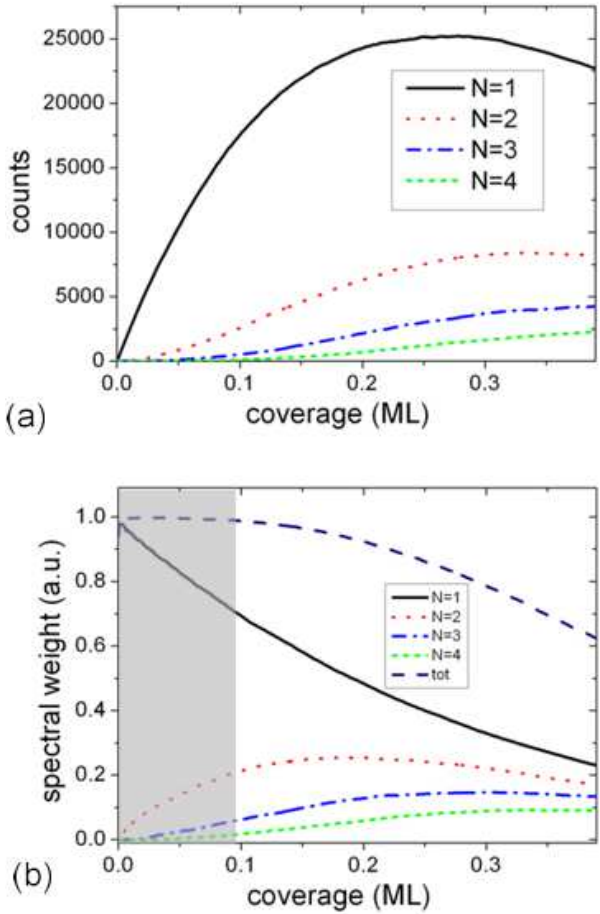}
\caption{(a) Distribution $\sum \limits_g \Gamma(N,g)$ of cluster sizes $N$ versus coverage and (b) spectral weights $\omega(N)$ of the different cluster sizes $N$. }
\label{statistics}
\end{figure}

After every MC deposition cycle the program counts the different types of clusters: an atom is evaluated as a monomer if it has no nearest neighbor (NN) on the same lattice, two atoms are evaluated as a dimer if they have just themselves as NNs, and so on. Moreover, the program distinguishes between different geometries $g$ for one and the same cluster size $N$, e.g. between linear trimers and trimers with an angle. The MC simulation thus gives access to the distribution $\Gamma(N,g)$ of all different cluster configurations ($N,g$) from monomers to tetramers ($N<5$) both on hcp or fcc sublattices. Figure~\ref{statistics}(a) reflects the statistics of cluster counts with size $N<5$ versus coverage.

Our DFT calculations show the importance of long-range RKKY interactions between pairs of monomers (see Fig.~\ref{exchange_J}),
which become important in the lowest coverage range. To capture these effects the statistics of single atom pairs is extracted, evaluating the combination of monomer pairs on hcp and fcc lattices from 2nd NN up to 5th NN distances for a given MC distribution. Herby, we only count pairs for which other monomers are found only at larger distances. This approximation thus assumes that for these pairs, residual oscillating RKKY field contributions of other surrounding monomers and pairs in average cancel each other and play a minor role.

Figure~\ref{statistics} illustrates the evolution of ($N,g$) with coverage. Figure~\ref{statistics}(a) shows the statistics of cluster counts with size $N$ versus coverage, while Fig.~\ref{statistics}(b) translates this statistics into spectral weights contributing to the X-ray absorption signal. The spectral weight $\omega(N)$ is hereby defined as:

\begin{equation}
\label{spectral weight}
{\omega(N)= \frac{\sum \limits_g N\cdot \Gamma(N,g)}{\sum \limits_{\tilde{N},\tilde{g}} \tilde{N}\cdot \Gamma(\tilde{N},\tilde{g})}}
\end{equation}

The degree of magnetic alignment of a given Fe cluster ($N$, $g$) (including RKKY-coupled pairs of monomers) with an applied field of $\bold{B}=B\bold\cdot {\hat{z}}$ is estimated using a Zeeman energy term of the form $E(M_{\text{tot}}^{N,g}, B, \Theta) = - B\cdot M_{\text{tot}}^{N,g} \cdot \text{cos}(\Theta)$, where the absolute value of the total moment vector $\bold{M}_{\text{tot}}^{N,g}$ in units $\mu_{\text{B}}$ is defined as the sum of total Fe moment $M_{\text{Fe}}$ and induced substrate moments $M_{\text{4d}}$:  $M_{\text{tot}}^{N,g} = M_{\text{Fe}} + M_{\text{4d}}$. $\Theta$ is the angle between the moment vector and the field direction ${\hat{z}}$. All moments are readily taken from DFT results. The contribution $R_{L_3}^{N,g}(B, T)$ of a certain cluster with $(N,g)$ to the total signal $R_{L_3}$ is then given by a Boltzmann statistics, allowing $M_{\text{tot}}^{N,g}$ to point in all directions in space:

\begin{widetext}

\begin{equation}
\label{Boltzmann}
{R_{L_3}^{N,g}(B, T) = {{R_{L_3}^{\text{sat}}}\over{N \cdot 3\mu_{\text{B}}}}{{\int \limits_0^{2\pi} \int \limits_0^{\pi} M_{\text{Fe}}^{N,g} \cdot \text{cos}(\Theta) \cdot \text{sin}(\Theta) \cdot e^{- E(M_{\text{tot}}^{N,g}, B, \Theta)/k_{\text{B}} T} d\Theta d\varphi} /Z}},
\end{equation}

\end{widetext}

where $Z$ is the partition function. The term $M_{\text{Fe}}^{N,g} \cdot \text{cos}(\Theta)$ projects the Fe cluster moments $M_{\text{Fe}}^{N,g}$ onto the ${\hat{z}}$ direction, which accounts for the fact that the XMCD technique measures Fe moment components along the X-ray beam direction. The calibration factor in front of the integral contains $R_{L_3}^{\text{sat}}=(0.6\pm 0.05)$, which is the value of $R_{L_3}$ expected for low coordinated Fe spin moments of $(3.0\pm 0.2)~\mu_{\text{B}}$ in a saturating magnetic field~\cite{5a}. In our simulation we thus make the assumption that the value $R_{L_3}^{\text{sat}}$ is a Fe moment dependent constant value which does not change significantly if the coordination state of the Fe changes. This assumption is not generally valid but is a good approximation for Fe atoms in metallic environments.
For a comparison with the experimental spectroscopy data, the total simulated signal $R_{L_3}$ is defined as the sum of all components $R_{L_3}^{N,g}(B, T)$ of clusters $(N,g)$ at experimental conditions $T=8$~K and $B=5$~T, scaled to their respective spectral weights determined by $\Gamma(N,g)$.

The coverage dependent spectral weight of clusters $N=1, 2, 3, 4$ given in Fig.~\ref{statistics}(b) shows that the coverage range up to $\theta=0.1$~ML is clearly dominated by monomers and dimers as expected (grey shaded region). At $\theta=0.2$~ML dimer contributions with $N=2$ are comparable to those of monomers. At the same time also clusters with $N=3$ gain a spectral weight of more than 10$\%$, indicating the onset of larger cluster contributions.

At coverages $\theta=0.25$~ML the spectral weight of all cluster contributions with $N>4$ reaches a value of 10$\%$.
In the framework of our simulations these contributions are considered to have zero average moment. As a consequence the simulation represents a lower limit of the expected signal especially for larger coverages. From the statistics we estimate the validity of our MC simulation to be limited to the coverage range $\theta<0.3$~ML, also because beyond this coverage we expect the onset of intermediate and 2nd layer formation, both not covered in our MC simulation.


\end{document}